# Observation of Fermi-surface-dependent nodeless superconducting gaps in $Ba_{0.6}K_{0.4}Fe_2As_2$

H. Ding[1], P. Richard[2], K. Nakayama[3], T. Sugawara[3], T. Arakane[3], Y. Sekiba[3], A. Takayama[3], S. Souma[2], T. Sato[3], T. Takahashi[2,3], Z. Wang[4], X. Dai[1], Z. Fang[1], G. F. Chen[1], J. L. Luo[1], and N. L. Wang[1]

[1] *Beijing National Laboratory for Condensed Matter Physics, and Institute of Physics, Chinese Academy of Sciences, Beijing 100080, China*
[2] *WPI Research Center, Advanced Institute for Material Research, Tohoku University, Sendai 980-8577, Japan*
[3] *Department of Physics, Tohoku University, Sendai 980-8578, Japan*
[4] *Department of Physics, Boston College, Chestnut Hill, MA 02467, USA*

**The recent discovery of superconductivity in iron-arsenic compounds below a transition temperature ($T_c$) as high as 55K [1,2,3,4,5] ended the monopoly of copper oxides (cuprates) in the family of high-$T_c$ superconductors. A critical issue in understanding this new superconductor, as in the case of cuprates, is the nature, in particular the symmetry and orbital dependence, of the superconducting gap. There are conflicting experimental results, mostly from indirect measurements of the low energy excitation gap, ranging from one gap [6] to two gaps [7,8], from line nodes [8] to nodeless[6,9] gap function in momentum space. Here we report a direct observation of the superconducting gap, including its momentum, temperature, and Fermi surface (FS) dependence in $Ba_{0.6}K_{0.4}Fe_2As_2$ ($T_c$ = 37 K) using angle-resolved photoelectron spectroscopy. We find two superconducting gaps with different values: a large gap ($\Delta \sim 12$ meV) on the two small hole-like and electron-like FS sheets, and a small gap ($\sim 6$ meV) on the large hole-like FS. Both gaps, closing simultaneously at the bulk $T_c$, are nodeless and nearly isotropic around their respective FS sheets. The isotropic pairing interactions are strongly orbital dependent, as the ratio $2\Delta/k_BT_c$ switches from weak to strong coupling on different bands. The same and surprisingly large superconducting gap due to strong pairing on the two small FS, which are connected by the ($\pi$, 0) spin-density-wave vector in the parent compound, strongly suggests that the pairing mechanism originates from the inter-band interactions between these two nested FS sheets.**

Our angle-resolved photoelectron spectroscopy (ARPES) in the normal state at $T$ = 50 K revealed three FS sheets in Ba$_{0.6}$K$_{0.4}$Fe$_2$As$_2$ single crystals. They consists an inner hole-like FS pocket (from now on we refer to it as the α FS) and an outer hole-like FS sheet (the β FS), both centered at the zone center Γ, and an electron-like FS (the γ FS) centered at M, or (π, 0) defined in the unreconstructed Brillouin zone (BZ). A more accurate FS contours can be traced out from an ARPES intensity plot near the Fermi energy ($E_F$) at low temperatures in the superconducting state, with the help of the sharp quasiparticle (QP) peaks that emerge below $T_c$ in this material as shown in Fig. 1**b**. The α FS is nearly circular with an enclosed area of ~ 4% of the reconstructed BZ area. The β FS is more square-like with an area of ~ 18%, and the electron-like γ FS is an ellipse elongated along the Γ-M direction occupying an area of ~ 3%. Thus, according to the Luttinger theorem, the total hole concentration is 38%, close to the nominal bulk hole concentration of 40% per two Fe atoms in a doubled unit cell. The observed FS topology (Fig. 3**a**) agrees well with the LDA band theory predictions at $k_z=\pi$. We caution that the $k_z$ dispersion of the β FS and a possible third hole-like pocket of similar size as the α FS, predicted by band calculations [10,11] may affect the carrier counting. When the samples are cooled down below $T_c$, as shown in the energy distribution curves (EDCs) in Figs. 1**d-m** at $T$ = 15 K, we clearly observe that the leading edge of the spectra on all three FS sheets shifts away from $E_F$, indicating the opening of an energy gap. In addition, sharp quasiparticle (QP) peaks are observed in the vicinity of the Fermi crossing ($k_F$) points, enabling us to estimate the gap values from the QP peak positions. As can be seen in Figs. 1**d-m**, the gap opened on the β FS is clearly smaller than those on the α and γ FS. It is also interesting to note that the spectral linewidth near the β FS is sharper than the ones near the α and γ FS, indicating a longer QP lifetime or reduced scattering rate for the low-energy β band excitations. In fact, the QPs near the α and γ FS have a similar unconventional QP lineshape, with an additional shoulder appearing on the low-energy side of the main QP peak, which will be discussed further below. Comparing to the normal state dispersion, we find that the QP dispersion (excluding the low-energy shoulder) in the superconducting state reaches a local minimum at $k_F$ with the bending-back effect, similar to the Bogoliubov QP (BQP) dispersion observed in the high-$T_c$ cuprates superconductors [12]. Such a dispersion behavior usually indicates the opening of a superconducting energy gap for the dispersive QPs.

To provide further evidence that the observed gap is indeed the superconducting gap, we have performed temperature ($T$) dependent measurements along the three cuts displayed in Fig. 2, crossing the three FS sheets respectively. Following a common practice in ARPES [13], we symmetrize the EDCs at $k_F$ to approximately remove the effect of the Fermi function to the leading edge and QP peak position, and extract the full gap (2Δ) from the separation of the two symmetrical QP peaks. Fig. 2**b** shows the $T$-dependence of a symmetrized EDC on the α FS. We clearly observe that the main QP peak position remains at a constant energy (~ 12 meV) from 7 K to 30 K and moves rapidly toward $E_F$ within the narrow temperature range of 35–40 K, accompanied by a sudden broadening of the linewidth. This is very similar to the behavior of a superconducting gap closing at $T_c$ observed in the overdoped high-$T_c$ cuprates [14]. The extracted gap values at different temperatures are plotted in Fig. 2**c**, which clearly shows that the gap collapses at the bulk $T_c$, at a rate steeper than the classic BCS curve (the red line in Fig. 2**c**). Similar $T$-dependence is also observed for the other two gaps on the β and γ FS sheets, as can be seen in Figs. 2**d-i**. In addition, the superconducting spectra can be recovered without any noticeable changes after several thermal cycles up to 50 K. The superconducting nature of these QP gaps are further supported by the observation of a residual BQP hole branch above $E_F$ [15], as shown in the inset of Fig. 2**d** for the β FS, where the full spectral function recovered after removing (dividing) the Fermi function shows good particle-hole symmetry, a hallmark associated with the superconducting gap.

We turn to the low-energy shoulder on the spectra of the α and γ FS, which seems to follow a different $T$-dependence; it broadens rapidly around 15–25 K, which is well below $T_c$, suggesting a different origin. At $T$ = 7 K, this shoulder evolves into a peak structure, but with a strongly sample dependent location generally within 6 meV of the Fermi level. In contrast, the superconducting gaps along the three FS sheets are robust and sample independent. Although further studies are needed to resolve this issue, we believe that the shoulder features are either due to surface effects or a minority phase that does not superconduct.

By utilizing the momentum-resolving capability of ARPES, we have mapped out the complete superconducting gaps along the three FS sheets. Symmetrized EDCs measured at $k_F$, some of which have been shown in Figs. 3**b-d**, are used to extract the *k*-dependence of the superconducting gap. It is apparent from viewing these EDCs that the superconducting gaps on the three FS sheets are nearly isotropic with a less-than-20% anisotropy. Fig. 3**e** displays the full *k*-dependence on a polar plot, confirming that nodeless superconducting gaps open on all the FS sheets. The *k*-averaged gap values are approximately 12, 6, and 12 meV for the α, β, and γ FS, yielding the ratio of $2\Delta/k_B T_c$ of 7.5, 3.7, and 7.5, respectively. While the ratio on the β FS is close to the BCS value (3.52), the larger ratio on the α and γ FS, similar to the value observed in many high-$T_c$ cuprates, suggests that pairing is in the strong coupling regime on these two FS sheets.

We summarize our main results in Fig. 4. We have demonstrated the multi-orbital nature of the superconducting state in single crystals $Ba_{0.6}K_{0.4}Fe_2As_2$, a prototypical hole-doped iron-arsenic superconductor. Nearly isotropic and nodeless superconducting gaps of different values open simultaneously at the bulk $T_c$ on all three observed FS sheets of electron and hole characters. The most natural interpretation of our findings is that the pairing order parameter has an *s*-wave symmetry, although we cannot rule out the possibility of nontrivial relative phases between the pairing order parameters on the different FS sheets. Perhaps the most striking feature is that strong pairing on the α and γ FS produced nearly the same, surprisingly large superconducting gaps. In the undoped parent compound, these two FS sheets are nested by the Q = (π, 0) spin-density-wave (SDW) wave vector[16]. In the sufficiently hole-doped superconducting sample, we have observed evidence of a similar band folding between Γ and M points at low temperatures which could be due to SDW fluctuations or short-range order by the same vector Q. Remarkably, the α and γ FS are still well connected by the Q-vector reminiscent of an inter-band nesting condition along large portions of the two FS. The latter can enhance the kinetic process where a zero momentum pair formed on the α (γ) FS is scattered onto the γ (α) FS by the fluctuations near the wave vector Q, whereby increasing the pairing amplitude. These observations strongly suggest that the inter-band interactions play an important role in the superconducting pairing mechanism of this new class of high-temperature superconductors.

**REFERENCES:**

[1] Kamihara, Y., Watanabe, T., Hirano, M. & Hosono, H. Iron-based layered superconductor La[$O_{1-x}F_x$]FeAs (x = 0.05–0.12) with $T_c$ = 26 K. J. Am. Chem. Soc. **130**, 3296–3297 (2008).

[2] Wen, H. H., Mu, G., Fang, L., Yang, H. & Zhu, X. Y. Superconductivity at 25 K in hole doped $La_{1-x}Sr_x$OFeAs. Europhys. Lett. **82**, 17009 (2008).

[3] Chen, X. H. *et al*. Superconductivity at 43 K in samarium-arsenide oxides $SmFeAsO_{1-x}F_x$. Preprint at (http://arXiv.org/abs/0803.3603) (2008).

[4] Chen, G. F. *et al*. Superconductivity at 41 K and its competition with spin-density-wave instability in layered $CeO_{1-x}F_x$FeAs. Phys. Rev. Lett. **100**, 247002 (2008).

[5] Ren, Z. A. *et al*. Superconductivity at 55 K in iron-based F-doped layered quaternary compound Sm[$O_{1-x}F_x$]FeAs. Chin. Phys. Lett. **25**, 2215 (2008).

[6] Chen, T. Y. *et al*. A BCS-like gap in the superconductor $SmFeAsO_{0.85}F_{0.15}$. Nature **453**, 1224-1227 (2008).

[7] Wang, Y. *et al*. Nodal superconductivity with multiple gaps in $SmFeAsO_{0.9}F_{0.1}$. Preprint at (http://arXiv.org/abs/0806.1986) (2008).

[8] Matano, K. *et al*. Spin-singlet superconductivity with multiple gaps in PrOFFeAs. Preprint at (http://arXiv.org/abs/0806.0249) (2008).

[9] Hashimoto, K. *et al*. Microwave penetration depth and quasiparticle conductivity in single crystal $PrFeAsO_{1-y}$: evidence for fully gapped superconductivity. Preprint at (http://arXiv.org/abs/0806.3149) (2008).

[10] Ma, F., Lu Z.-Y. & Xiang, X., Electronic band structure of $BaFe_2As_2$. Preprint at (http://arXiv.org/abs/0806.3526) (2008).

[11] Xu, G., Zhang, H., Dai, X. & Fang, Z. Electron-hole asymmetry and quantum critical point in hole-doped FeAs-based superconductors. To be published.

[12] Campuzano, J. C. *et al*. Direct observation of particle-hole mixing in the superconducting state by angle-resolved photoemission, Phys. Rev. B **53**, R14737 (1996).

[13] Norman, M. R. *et al*. Destruction of the Fermi surface in underdoped high $T_c$ superconductors. Nature **392**, 157 (1998).

[14] Norman, M. R., Randeria, M., Ding, H. & Campuzano, J. C., Phenomenology of photoemission lineshapes of high $T_c$ superconductors. Phys. Rev. B **57**, R11093 (1998).

[15] Matsui, H. *et al*. BCS-Like Bogoliubov quasiparticles in high-$T_c$ superconductors observed by angle-resolved photoemission spectroscopy. Phys. Rev. Lett. **90**, 217002 (2003).

[16] Huang, Q. *et al*. Magnetic order in $BaFe_2As_2$, the parent compound of the FeAs based superconductors in a new structural family. Preprint at (http://arXiv.org/abs/0806.2776) (2008)

[17] Chen, G. F. *et al*. Breaking rotation symmetry in single crystal $SrFe_2As_2$. Preprint at (http://arXiv.org/abs/0806.2648) (2008).

## ACKNOWLEDGEMENTS

This work was supported by grants from Chinese Academy of Sciences, NSF, Ministry of Science and Technology of China, JSPS, JST-CREST, MEXT of Japan, and NSF, DOE of US.

Correspondence and requests for materials should be addressed to H.D. (dingh@aphy.iphy.ac.cn)

## FIGURE CAPTIONS

FIG. 1(COLOR)

**Fermi surface and energy band dispersion of $Ba_{0.6}K_{0.4}Fe_2As_2$.**

High-quality single crystals of $Ba_{0.6}K_{0.4}Fe_2As_2$ used in our study were grown by the flux method [17]. High-resolution ARPES measurements were performed in the photoemission laboratory of Tohoku University using a microwave-driven Helium source ($h\nu$ = 21.218 eV) with an energy resolution of 2-4 meV, and momentum resolution of 0.007 Å$^{-1}$. Samples were cleaved *in situ* at 15 K and measured at 7-50 K in a working vacuum better than $5 \times 10^{-11}$ Torr. Low-energy electron diffraction on a measured surface shows a sharp 1x1 pattern without any detectable reconstruction down to 80K. Mirror-like sample surface was found to be stable without obvious degradation for the measurement period of 3 days. **a**, AC susceptibility of $Ba_{0.6}K_{0.4}Fe_2As_2$ ($T_c \sim 37$ K) as a function of temperature. **b** Representative ARPES spectra in the vicinity of $k_F$ measured in the superconducting state ($T$ = 15 K) at two points marked by the circles in the BZ shown in panel c**.** Clear dispersion and sharp QP peaks were observed, indicating good sample (surface) quality. **c** FS contour determined by plotting the ARPES spectral intensity integrated within ±10 meV with respect to $E_F$. **d-h**, ARPES spectral intensity at 15 K as a function of wave vector and binding energy and the corresponding EDCs (**i-m**) measured along several cuts in the BZ shown in panel c.

FIG. 2(COLOR)

**Direct observation of multiple superconducting gaps.**

**a**, *T*-dependence of EDC measured at the $k_F$ point of the α FS (red dot in inset). **b,** the symmetrized EDC. Dashed line in **b** denotes the position of the quasiparticle peak. **c**, *T*- dependence of the superconducting energy gap size. Solid line is the BCS mean-field gap equation with $T_c = 37$ K and zero temperature gap $\Delta_\alpha(0) = 12.5$ meV. **d-f**, (**g-i**) same as **a-c**, but measured on the $k_F$ point of the β (γ)FS (blue (green) dots in the inset to **a**). The zero temperature gap value used in the BCS equation (solid line in **f**) is $\Delta_\beta(0) = 5.5$meV. Note that the rapid drop of the gap value on approaching $T_c$ from the superconducting side cannot be described by the BCS mean-field theory. Inset in **d** shows the expansion of the EDC near $E_F$ at 25 K (red), together with the EDC (blue) after dividing out the Fermi function. Nearly particle-hole symmetric BQP peaks are clearly visible, indicating the superconducting nature of the energy gap.

FIG. 3(COLOR)
**Momentum and orbital dependence of the superconducting gap.**
**a**, **b**, **c**, Symmetrized EDCs at 15 K measured at various $k_F$ points on the α-, β-, and γ-FS, labeled by respective colored symbols correspondingly. **d,** Extracted FS from ARPES measurements in the superconducting state. **e**, Superconducting gap values at 15 K extracted from the EDCs (**a**, **b**, and **c**) shown on polar plot for the α, β– (left) and γ– (right) FS as a function of FS angle ($\theta$) (zero degree is along Γ-M). Nearly isotropic superconducting gap with weak anisotropy can be seen for all three observed FS sheets.

FIG. 4(COLOR)
**Multi-orbital nature and FS-dependent nodeless superconducting gaps in $Ba_{0.6}K_{0.4}Fe_2As_2$**
Three-dimensional plot of the superconducting-gap size ($\Delta$) measured at 15 K on the three observed FS sheets (shown at the bottom as an intensity plot) and their temperature evolutions (inset).

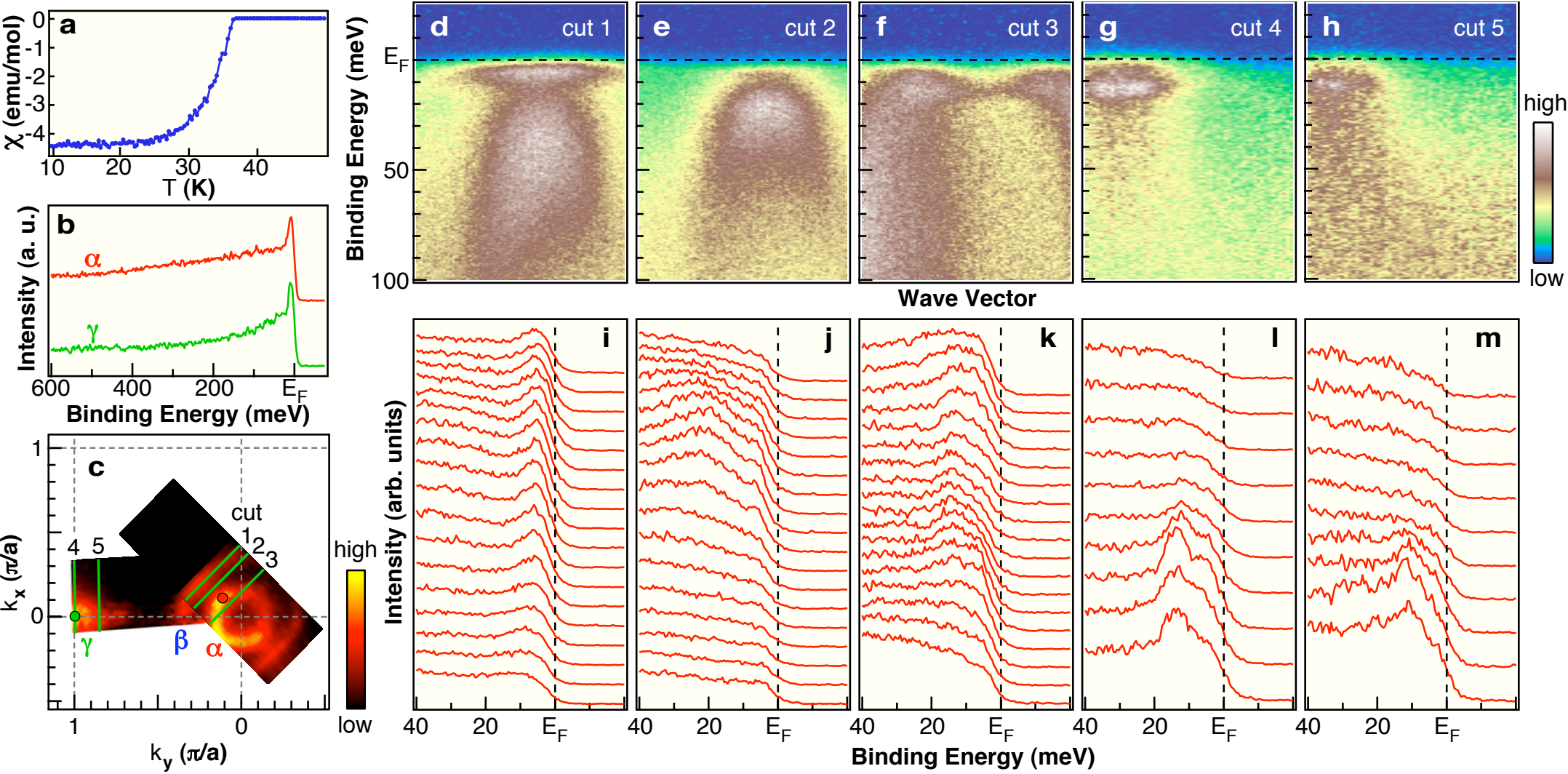

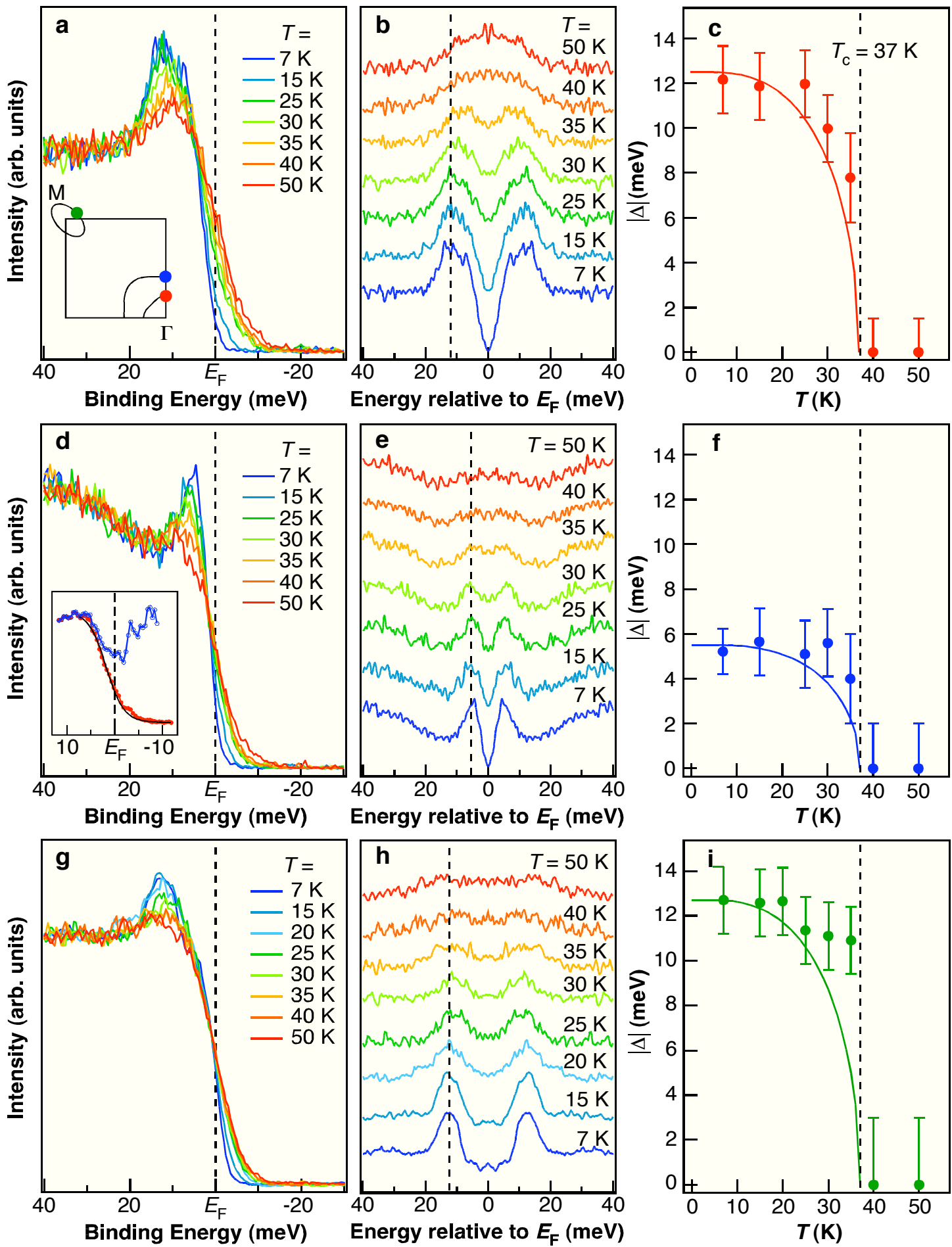

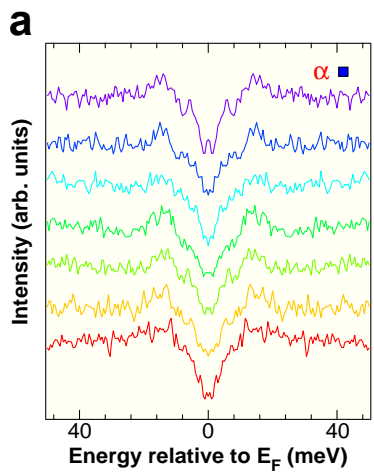 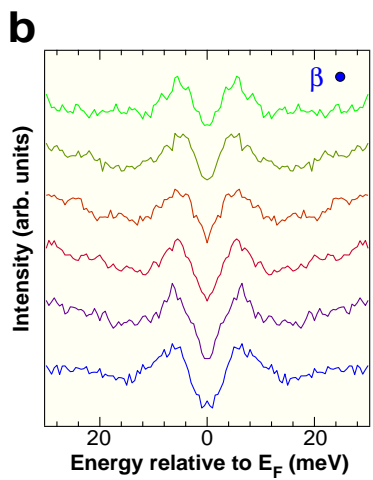 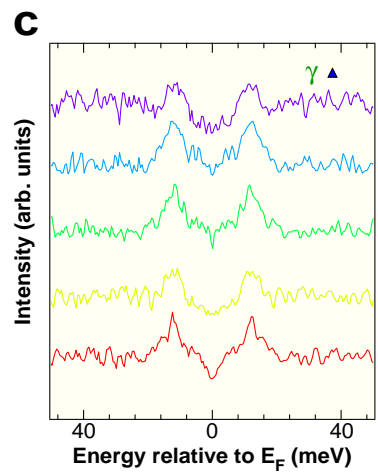
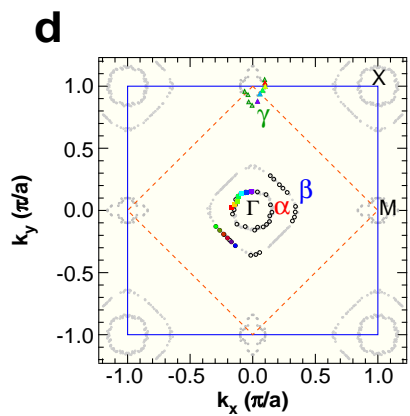 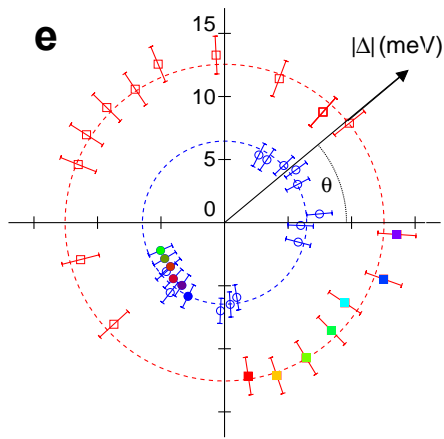 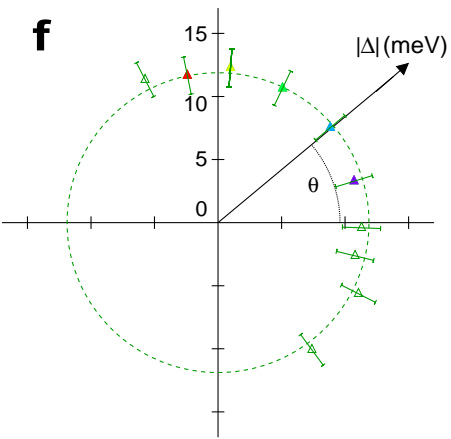

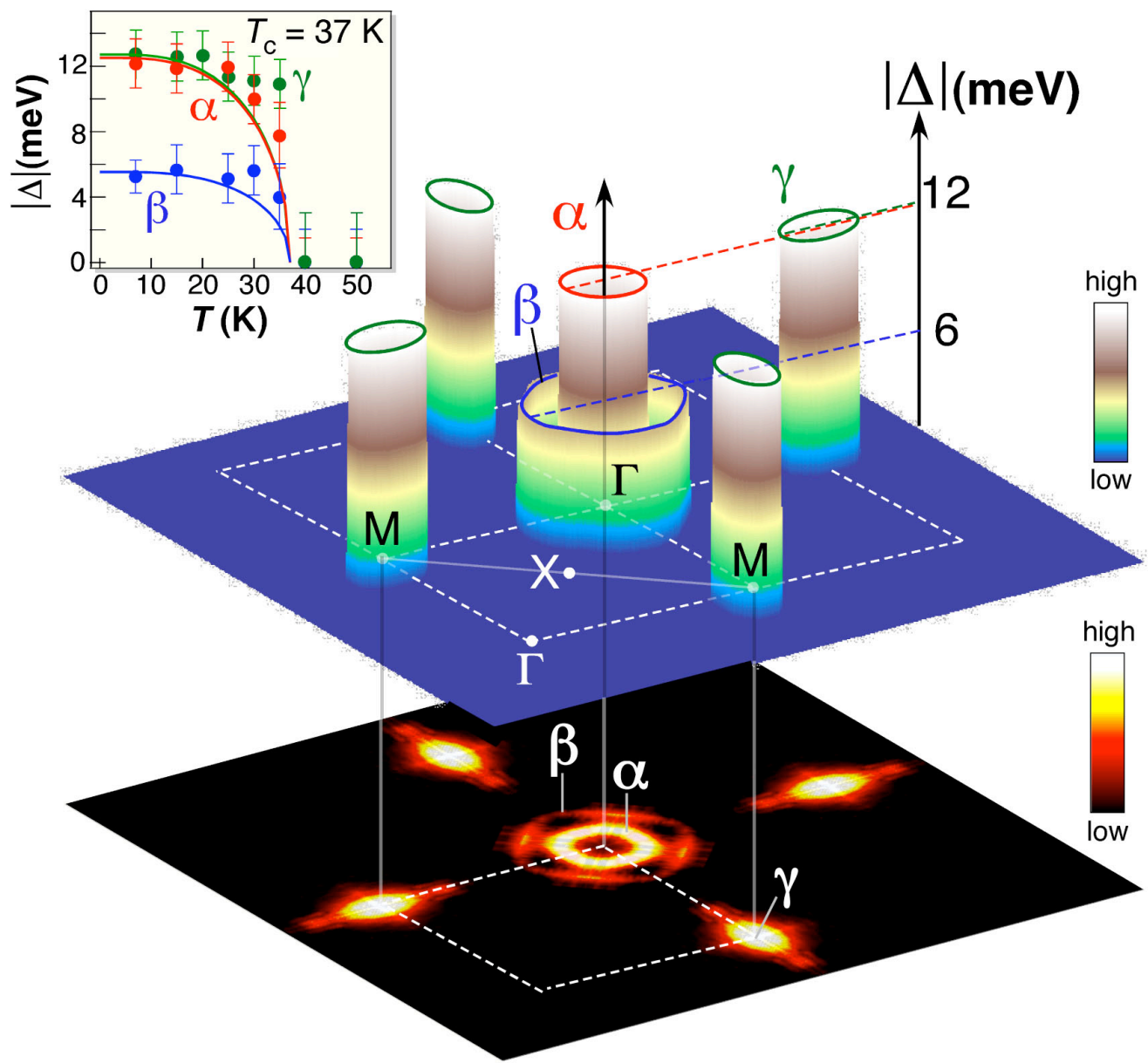